\documentclass[twocolumn, aps, prl, 10pt]{revtex4}

\usepackage{graphicx}
\usepackage{amsmath,amsfonts}
\usepackage{lipsum}

\usepackage{cleveref}
\usepackage{float}
\usepackage{xfrac}
\usepackage[usenames, dvipsnames]{color}
\usepackage{cancel}
\usepackage{xcolor}
\usepackage{array}

\usepackage{tabu}
\usepackage{rotating}

\newcolumntype{L}[1]{>{\raggedright\let\newline\\\arraybackslash\hspace{0pt}}m{#1}}
\newcolumntype{C}[1]{>{\centering\let\newline\\\arraybackslash\hspace{0pt}}m{#1}}
\newcolumntype{R}[1]{>{\raggedleft\let\newline\\\arraybackslash\hspace{0pt}}m{#1}}

\begin{document}

\title{Attosecond-Resolution Hong-Ou-Mandel Interferometry}

\author{Ashley Lyons$^{1}$, George~C.~Knee$^{2}$, Eliot Bolduc$^{1}$, Thomas Roger$^{1}$, Jonathan Leach$^{1}$, Erik~M.~Gauger$^{1}$, Daniele Faccio$^{1}$}

\affiliation{$^{1}$School of Engineering and Physical Sciences, Heriot-Watt University, Edinburgh, EH14 4AS, UK}
\affiliation{$^{2}$Department of Physics, University of Warwick, Coventry, CV4 7AL, UK.}

\date{\today}

\begin{abstract}{
When two indistinguishable photons are each incident on separate input ports of a beamsplitter they `bunch' deterministically, exiting via the same port as a direct consequence of their bosonic nature. This two-photon interference effect has long-held the potential for application in precision measurement of time delays, such as those induced by transparent specimens with unknown thickness profiles. However, the technique has never achieved resolutions significantly better than the few femtosecond (micron)-scale other than in a common-path geometry that severely limits applications. Here we develop the precision of HOM interferometry towards the ultimate limits dictated by statistical estimation theory, achieving few-attosecond (or nanometre path length) scale resolutions in a dual-arm geometry, thus providing access to length scales pertinent to cell biology and mono-atomic layer 2D materials.}
\end{abstract} 

\maketitle


Since its discovery, Hong-Ou-Mandel (HOM) interferometry \cite{Hong1987} has found a wide variety of applications within quantum optics \cite{Ou_book, Shih1988,Rarity1990,Walther2004,Kok2007,Ma2012}. For example, it is commonly exploited as a measure of the distinguishability of photons produced by quantum dots \cite{Santori2002, Kim2016a}. It can be used as as a source of two photon N00N states: a class of states widely studied in quantum metrology owing to their ability to reach the Heisenberg limit in phase sensitive measurements \cite{Boto2000,Lee2002,Dowling2008,Rozema2014,Sidhu2016}. HOM interferometry is impervious to changes in the relative phase between the two photons, a property which implies that a HOM based sensor does not require potentially impractical or expensive stabilisation, as is typically required in classical interferometry.

To date, the highest precision time-delay measurements employing the HOM effect make use of orthogonally polarised photon pairs to measure polarisation mode dispersion \cite{Branning2000, Dauler2000}. These studies have produced measurements of the group delay between pairs propagating along a common path, to within a 0.1 fs uncertainty. The common-path geometry significantly aids the stability of the interferometer, but can only be applied to (and thus is only relevant for) birefringent samples. A much wider range of applications is possible if the same or better precision can be achieved with a dual-arm geometry which allows for a delay to be introduced by an arbitrary means.

Closely related to HOM interferometry, Quantum Optical Coherence Tomography (QOCT) is a method for extracting depth profiles of reflective interfaces, also via a HOM measurement of the relative delay between photon pairs, often implemented in a dual-arm geometry. In this context, features on the order of a micron have been detected, including those introduced by biological specimens \cite{Abouraddy2002,Nasr2004,Lopez-Mago2012,Mazurek2013}. The limited depth resolution makes these approaches inadequate for smaller biological samples such as cell membranes, DNA samples or protein monolayers which have thicknesses on the order of 1 - 10 nm. 

Both QOCT and standard HOM measurements have to-date relied on detecting the shift in the interference minimum in the coincidence counts between the output ports of the interferometer. 

Here we devise and implement a completely new measurement and estimation strategy based on a Fisher information analysis. By tuning the interferometer to the delay that contains the maximum information content, and then by employing a maximum-likelihood estimation procedure, we achieve an improvement in precision and accuracy by two orders of magnitude over previous approaches \cite{prec_note}. We conducted measurements of the change in relative arrival time between two photons, $\delta \tau$, with an average accuracy of 6 as (1.7 nm) and average precision of 16 as (4.8 nm). Our best achieved accuracy was 0.5 as (0.15 nm) and best achieved precision was 4.7 as (0.9 nm). HOM interferometry can therefore enable single photon characterisation of optically transparent samples with thicknesses and length scales relevant for example to cell biology.

\begin{figure*}
\centering
\includegraphics[width = \textwidth]{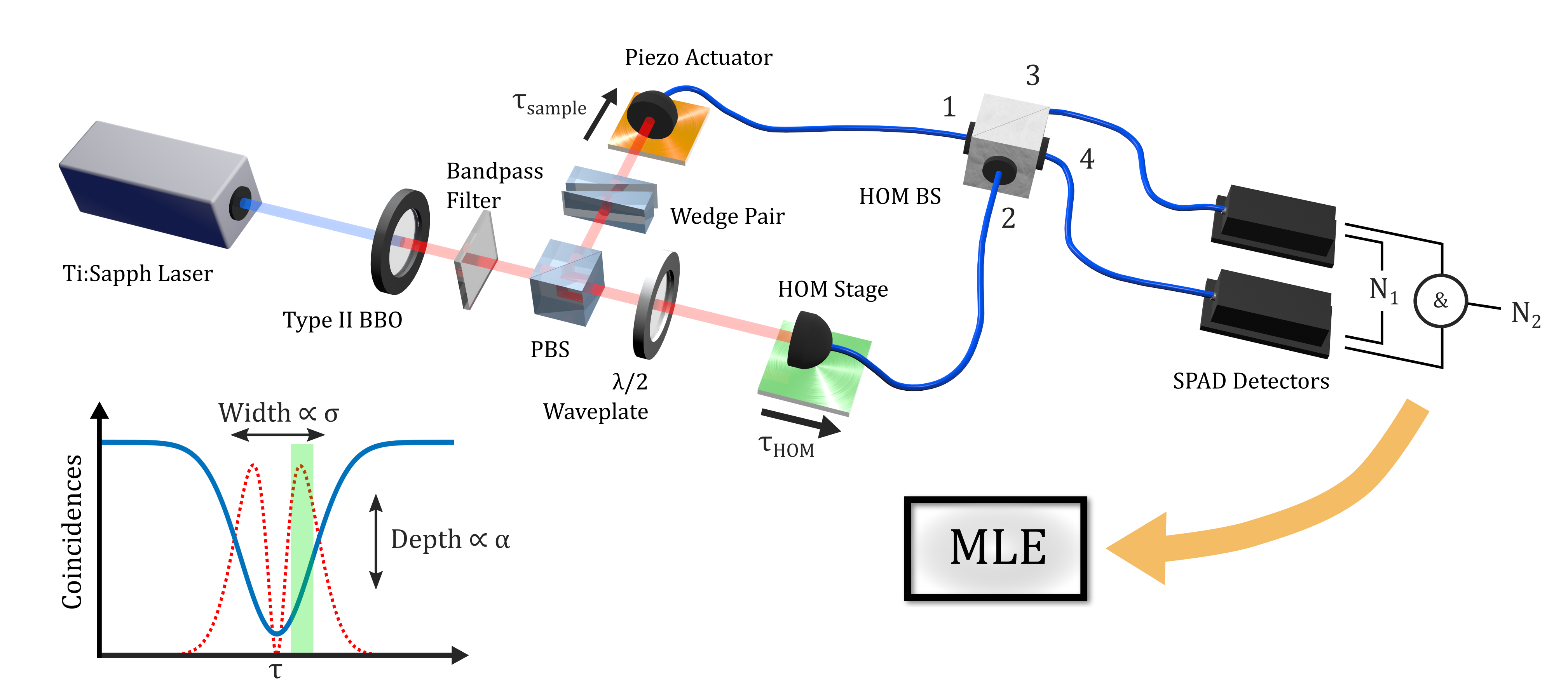}
\caption{A dual arm HOM interferometer. A pumped Type II BBO crystal is used as a source of SPDC photon pairs which are separated by a PBS before being subject to a differential time delay and recombination on separate input ports (1 and 2) of a fibre-coupled 50:50 beamsplitter (HOM BS). Coarse control of the optical delay  is achieved by a motorised translation stage (HOM Stage, controlling $\tau_\text{HOM}$) whilst fine control is achieved using a piezo actuator (controlling $\tau_\text{sample}$ and representing a transparent sample). From a scan of the HOM dip (Bottom left - blue) a peak in the Fisher information (red) is identified to be used in the sensing procedure (green). The difference in temporal delay between the two photons ($\tau := \tau_\text{sample} - \tau_\text{HOM}$), is quantified through Maximum Likelihood Estimation (MLE).}
\label{setup_fig}
\end{figure*}
%
%

If two photons are incident on the input ports of a balanced beamsplitter (BS), the probability of a coincident detection at the output ports depends on the inner product of the quantum states of each photon, and is influenced by the difference in arrival times of the two photons $\tau$. HOM interference is characterised by the coincidence rate falling to zero as all distinguishing information is erased: the so called `HOM dip' (see Figure~\ref{setup_fig}).
As Hong, Ou and Mandel showed in their original work, if $\tau$ is scanned, the minimum position of the interference pattern can be measured with at least sub-picosecond precision. Similar techniques were used in later works~\cite{Branning2000,Giovannini2015} and typically involve a simple least-squares fitting procedure for a scan over $\tau$. By contrast here we use Fisher information analysis as the key theoretical tool for unlocking a peak performance HOM interferometer. This allows us to introduce measurement and estimation protocols that are optimised to give a higher precision for a set amount of time invested or, equivalently, a greater information gain per photon. 

The ultimate limit on the precision of estimation is known as the Cram\'er-Rao bound \cite{Kok_book}, which states that the variance of any unbiased estimator (i.e. one whose expectation is equal to the true value of the parameter - see \emph{Bias} section in Appendix A) must be bounded by
\begin{align}
\text{Var}(\tilde{\tau})\geq 
\frac{1}{NF} ~,
\label{CRB}
\end{align}
where $\tilde{\tau}$ denotes an estimator for the parameter $\tau$. The Fisher information $F$ measures the amount of information about $\tau$ that can be extracted from a particular experiment. 

The HOM interferometer is characterised by many desirable features: for example, the large dynamic range (see Supplementary Information). In this work, however, we are primarily concerned with minimising $\text{Var}(\tilde{\tau})$. We achieve attosecond precision by accomplishing a combination of three goals: (i) maximising $F$,  (ii) saturating inequality (\ref{CRB}), and  (iii) increasing $N$ (i.e.~the number of repetitions of the experiment) as much as possible within the confines of slow drift in the setup.  

\begin{figure*}
\centering
\includegraphics[width = \textwidth]{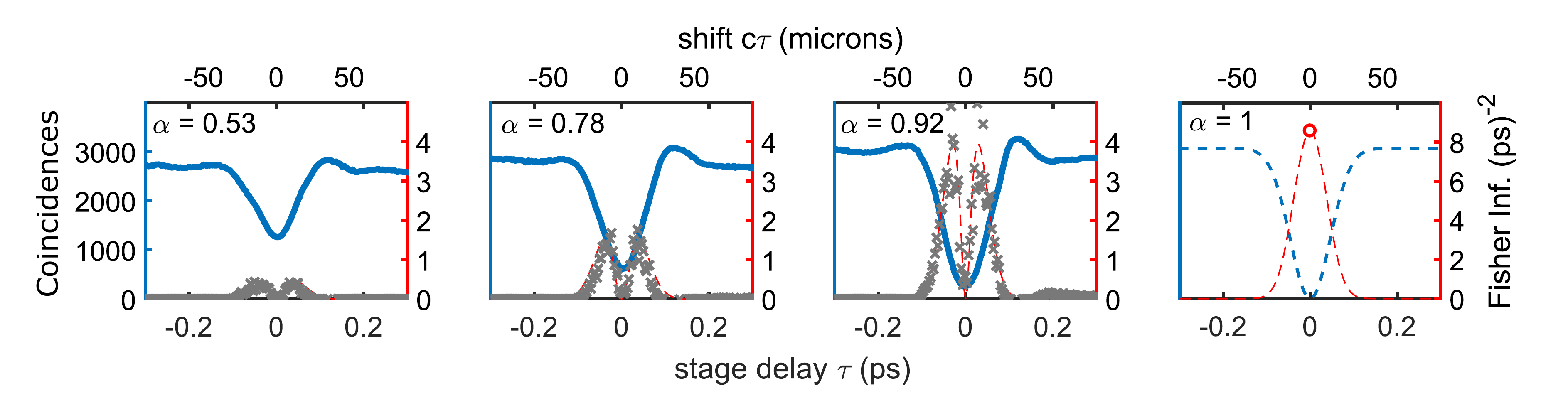}
\caption{Experimental HOM dips (left axis) are shown for various visibilities, introduced by a differential polarisation change.  Also shown is the estimated total Fisher information $NF$ (equation (\ref{Fisher}), red, right axis) along with the inverse-variance of our experimental estimates (grey crosses, right axis). The rightmost panel includes theoretical curves for perfect visibility, where the two peaks in the Fisher information have asymptotically merged. The open circle denotes a point where the Fisher information is undefined.}
\label{FI_plots_fig}
\end{figure*}

To achieve the first of our goals, we need to consider the dependence of $F$ on $\tau$ and other parameters.  Our statistical model is defined by a set of probabilities $P_{i}$, where $i=0,1,2$ denotes the number of detectors that click in each run. The probabilities depend on the following parameters: the wave-packet duration $\sigma$, which is proportional to the FWHM of the temporal mode function of each photon (here taken to be Gaussian functions); the maximum indistinguishability $\alpha$ (which sets the visibility of the interference); and the photon loss rate $\gamma$ (see Appendix A for the full model). The Fisher information evaluates to
%
\begin{align}
F=&\frac{1}{\sigma^2}\left(\frac{2\alpha ^2 \tau^2e^{-2\tau^2/\sigma^2}}{\frac{(1-\alpha e^{-\tau^2/\sigma^2})}{(1-\gamma)^2}-\frac{1}{2}(1-\alpha e^{-\tau^2/\sigma^2})^2}\right). 
\label{Fisher}
\end{align}
The distribution of information Eq.~(\ref{Fisher}) is doubly-peaked and symmetric around $\tau =0$  (see red dashed curves in Figures~\ref{setup_fig} and~\ref{FI_plots_fig}). We note that when $\alpha\rightarrow1$ (perfect visibility) the peaks asymptotically merge at the origin. As the visibility is reduced (as it is in all experiments) the maximum information decreases and the peaks move outwards. A similar phenomenon was observed in \cite{Jachura2016}, where non-unit visibility caused a dramatic qualitative change in the distribution of Fisher information for a Mach Zehnder interferometer. Here the changing distribution reflects the competition between i) the derivative of the inverted-Gaussian dip (which is optimised at $\tau = \sigma/\sqrt{2}$) and ii) the variance of $P_i$ (which is optimised at $\tau=0$). 
This suggests that using prior knowledge of $\tau$ and $\alpha$ enables the interferometer to be tuned to operate at the optimum point between these extremes. In the Appendix we discuss how we found this point.  

To achieve our second goal of saturating the Cram\`er-Rao bound, we employ the maximum likelihood estimator,
\begin{align}
\tilde{\tau}=\pm\sigma\sqrt{\ln \left(\frac{\alpha  (N_1+N_2)}{ N_1 -  N_2(\frac{1+3\gamma}{1-\gamma })}\right)}. 
\end{align}
where $N_1$ (respectively $N_2)$ is the number of times only one (both) detector(s) clicked (for details see the Appendix). This estimator is efficient, i.e. it saturates Eq.~(\ref{CRB}) when the number of trials is large enough \cite{Wolfowitz1965}. When the argument of the logarithm is negative, the likelihood is maximised at $\tau\rightarrow\pm\infty$. To account for this situation (which arises when the dataset is very noisy) we use a Bayesian analysis (see Appendix). The estimator is non-linear but analytically calculable and thus suitable for real-time estimation with the data. Note that due to the symmetry of the HOM dip there is a two-fold ambiguity in the estimate -- we can only obtain its magnitude and not its sign. As we shall see later this issue will be resolved through our measurement protocol. 
Note that the estimation involves observable quantities $N_i$ as well as $\gamma,\sigma$ and $\alpha$. The latter need to be separately estimated before measurements begin (see Appendix). 



To test our theory and demonstrate our protocol, a non-common path HOM interferometer was constructed using SPDC from a Type II nonlinear crystal as a source of orthogonally polarised photon pairs (see Figure~\ref{setup_fig} and Appendix). The photons are deterministically separated via their polarisation and each is collected by a single-mode fiber coupled to a beamsplitter (HOM BS, Figure~\ref{setup_fig}). The relative path length difference between the photon pair was controlled with a coarse stage.

The HOM dip was scanned with a 10 nm bandpass filter positioned before the PBS to ensure spectral indistinguishability and therefore high visibility two photon interference. 
As a first step, a total of 50 scans of the HOM dip were acquired in order to ascertain the variance of the estimates and compare with the predicted Fisher information (which acts as an upper bound on the inverse variance). In order to test the theoretical model further, the polarisation of one of the arms of the interferometer is rotated so as to reduce the visibility of the interference to approximately 50\%. The inverse variance of our estimates follows the predicted Fisher information distribution, as shown in Figure~\ref{FI_plots_fig}.

These data therefore constitute a good confirmation of Eqs.~(\ref{CRB})  and (\ref{Fisher}). By replacing the interference filter it was observed that both the measurement accuracy and precision were improved for wider bandwidth photons: this is despite of the reduction in $\alpha$, and due to the decrease in $\sigma$ and increase in $N$. For this reason, the bandpass filter was removed and replaced with a long-pass edge filter to block the pump field of the SPDC without altering the spectrum of signal and idler. 


 
For subsequent measurements, we introduced attosecond-scale temporal delays, $\tau_\text{sample}$ (see Figure~\ref{setup_fig}), with an additional piezo actuator (thus playing the role of a transparent sample that can be inserted in and out of the photon path). The HOM interferometer delay ($\tau_\text{HOM}$) was first tuned with a coarse control stage to a maximum in the Fisher information (see Figure~\ref{FI_plots_fig}) and was kept there whilst a large amount of data [$O(10^{9})$ counts] was collected. 

To achieve our third goal (increasing the number of measurements within the limits of experimental drift), the piezo actuator was periodically switched (every 100 ms) between the two positions (which we label {`in'} and {`out'}), and we collected a set of counts $N_i$ corresponding to each. We then use this data (in combination with the parameters $N_i$, $\gamma$, $\sigma$ and $\alpha$) to estimate $\tau^{\text{in}}$ and $\tau^{\text{out}}$. Finally, we extract the difference in differential time delay $\delta \tau := \tau^{\text{in}} -  \tau^{\text{out}} = (\tau_\text{sample}^\text{in} - \tau_\text{HOM}) - (\tau_\text{sample}^\text{out} - \tau_\text{HOM}) = \tau_\text{sample}^\text{in} - \tau_\text{sample}^\text{out}$.

\begin{figure*}
\centering
\includegraphics[width = \textwidth]{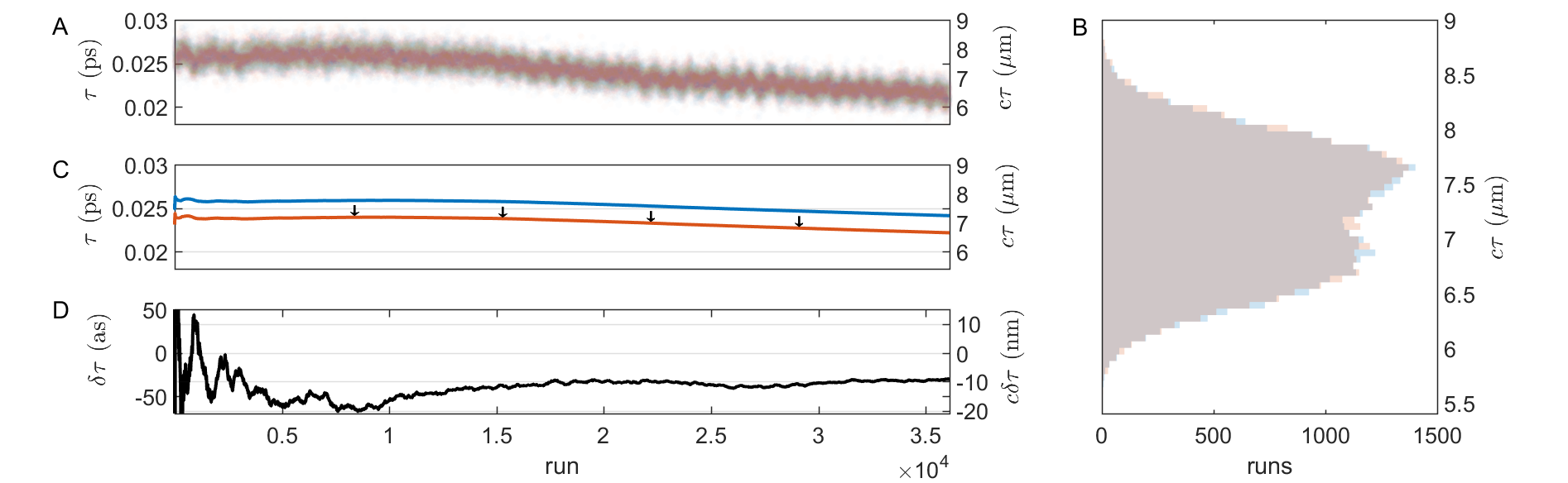}
\caption{A) Individual estimates of $\tau^{\text{in}}$ (blue) and $\tau^{\text{out}}$ (red) for two piezo positions separated by 10 nm (33.3 as). B) The histogram shows two, almost perfectly overlapping distributions. The distributions are generally non-unimodal, which is indicative of significant drift (or slowly varying noise). C) Cumulative estimates are plotted. The drift in each estimate is considerable: being approx. 2 fs (600 nm). The red curve has been shifted down by 2 fs for clarity (shown by the arrows). The drift for each sample position is very well correlated, because we switch the sample position much faster than the drift. D) Because of this correlation, the difference in cumulative estimates $\delta \tau$ is very stable and converges on the true value. }
\label{runs_fig}
\end{figure*}

\Cref{runs_fig} shows data for a sample position separation ($c \delta \tau$) of 10 nm. Each individual integration window yields a relatively poor precision $\sqrt{\textrm{Var}(\tilde{\delta} \tau)}=\sqrt{\textrm{Var}(\tilde{\tau}^{\text{in}})+\textrm{Var}(\tilde{\tau}^{\text{out}})}\approx600$ as (180 nm) estimate of the optical delay. By assuming each experiment is independent, we can combine $M=O(10^4$) such datasets to achieve a $O(100)$ fold improvement in precision, i.e towards a few attoseconds (few nm). The precision obtained using the entire dataset is estimated as $\sqrt{\textrm{Var}(\delta\tilde{\tau})/M}$.

%

\begin{figure}[h]
\centering
\includegraphics[width=8.5cm]{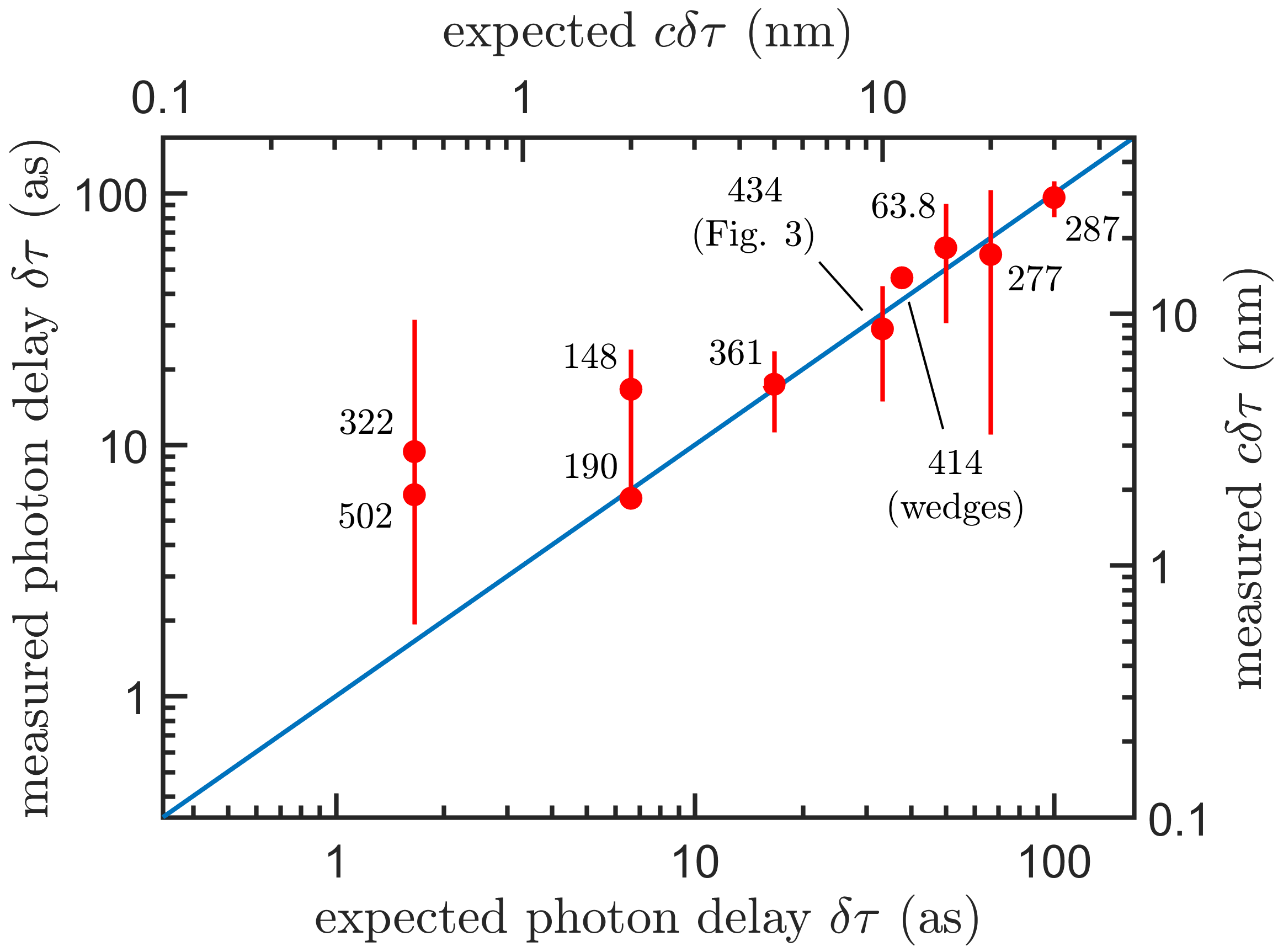}
\caption{Experiments: measured photon delays induced by the piezo shown against the set values on the piezo actuator. Number of individual measurements and integrations times vary as indicated in the plot (labels denote billions of incident biphotons). Error bars represent an interval of length $2\sqrt{\textrm{Var}(\tilde{t})}$. The datapoint corresponding to the glass wedges should only be read on the top and right axes (because of the nonunit refractive index).}
\label{deviation_fig}
\end{figure}

Next, a series of target piezo displacements ($\tau_1$) were set to test the capabilities of our protocol. \Cref{deviation_fig} shows the final estimated shifts compared to the ground truth displacements recorded by the internal capacitive sensor of the piezo actuator. The measurement procedure consistently returns a high degree of accuracy even down to the smallest set displacements of 1.5 as (0.5 nm): typical values of the measurement precision are within the $\pm6-15$ as ($\pm$2 - 5 nm) range.


We performed a final experiment to demonstrate the potential for our scheme to measure samples scanned transversely across the photon path and thus perform full imaging tasks. We introduce a controlled delay using a pair of transparent wedges positioned in one of the interferometer arms resulting in an asymmetric loss of around 30\%. The wedges are arranged such that translating one of the wedges changes the length of propagation through the glass whilst maintaining the alignment of the system (as shown in \Cref{setup_fig}). A target delay of 57 as, resulting from an estimated 11 nm of additional glass, was introduced using the wedge pair (taking the refractive index of glass to be 1.5). Our measurement procedure returned a measured delay of $69 \pm 5$ as giving an estimate of $14 \pm 1$ nm for the additional glass length. We attribute the difference between the measured and expected values to an imperfect calibration of the wedge system.

If we compare our results to the literature, the best HOM measurement performed to date had an accuracy of 200 as (60 nm) and a precision of 100 as (30 nm) \cite{Branning2000} utilising a common-path interferometer geometry. We show an accuracy improvement of approximately $31\times$ with neither the benefit of the inherent stability nor the limitation to birefringent samples. If compared to non-common path HOM interferometers, previous work showed resolutions of order of a few fs ($\sim \mu$m), with respect to which we have a more than 100$\times$ improvement \cite{Giovannini2015,Abouraddy2002,Nasr2004,Lopez-Mago2012,Mazurek2013}. Most importantly, our technique opens for the first time the possibility of using HOM interference to perform measurements of transparent samples in the single-attosecond delay (i.e. sub-nm path length) regime. Scan-free imaging capability could also be potentially introduced by resorting to wide-field lensless approaches recently demonstrated with a classical interferometer \cite{Terborg2016}. With a small modification the technique can also be applied to reflective samples such as those used in QOCT experiments providing the same enhancement to the precision. Furthermore, there is significant scope to increase the precision of our experiment yet further through the use of shorter down-conversion crystals (lower $\sigma$) and/or higher efficiency photodetectors (lower $\gamma$) \cite{Dauler2000}. For example, combining our method with the engineered source of photon pairs from the work of Okano et al. has the potential to yield another 30-fold improvement in precision arising from the inreased bandwidth of the photon pairs \cite{Okano2015}.
The HOM dip can also be specifically tailored to further optimise the amount of Fisher information obtainable. By using the best available photodetectors with upwards of 95\% efficiency ($\gamma = 0.05$) \cite{Marsili2013} and using Eq.~\ref{Fisher} we estimate that (holding $\alpha$ constant at 0.9) one could achieve an approximate 50$\times$ improvement in the Fisher information, or around 7$\times$ improvement in precision. 
\\
Finally, we note that our interferometer is capable of producing phase sensitive fringes (as shown in the Appendix), by rotating the PBS away from being perfectly aligned with the signal and idler polarisation reference frame. Here these fringes, that are due to single photon interference, have been suppressed in order to investigate the attainable precision using two-photon interference alone. In future, however, it would be possible to further increase the Fisher information by simply rotating the input photon polarisation (see Appendix E for details). This would allow for a further 150 fold improvement in precision, allowing measurements to reach into the picometer length scale.

\section*{Acknowledgments}
GCK thanks Jes\'us Rubio for helpful discussions. EMG acknowledges support from the Royal Society of Edinburgh and the Scottish Government. GCK was supported by the Royal Commission for the Exhibition of 1851. D.F. acknowledges support from the European Research Council under the European Union’s Seventh Framework Programme (FP/2007-2013)/ERC, Grant No. GA 306559, the Engineering and Physical Sciences Research Council (EPSRC, UK, Grants No. EP/M006514/1 and No. EP/M01326X/1) and the Leverhulme Trust.

\bibliographystyle{unsrtnat_edit}

\clearpage
\onecolumngrid
\normalsize
\appendix
\section*{Appendix A: Theory}
\begin{figure}[h]
\centering
\includegraphics[width=0.45\textwidth]{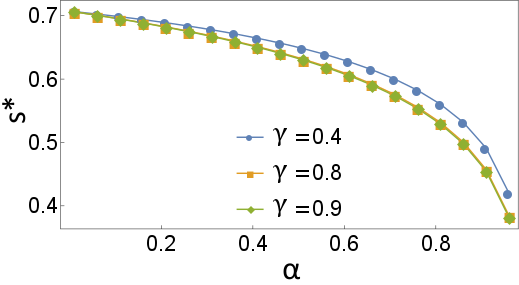}
\caption{Optimal sensitivity point as a function of visibility. The three curves correspond to different values of the photon loss rate $\gamma$.}
\label{ystar}
\end{figure}

{\emph{Quantum mechanical derivation of the HOM effect:}} 
Assume we have a source that can produce two-photon states of the form
\begin{align}
|\psi\rangle = \left(\eta[a_1^\dagger a_2^\dagger]+\sqrt{1-|\eta|^2}[b_1^\dagger a_2^\dagger]\right)|\text{vac}\rangle ,
\end{align}
where the operator $a_i^\dagger$ creates a photon with certain properties (frequency distribution, polarisation and so on) in mode $i=1,2$ corresponding to the two input modes of a balanced beam splitter (BS) (see Figure~\ref{setup_fig}), $b_i^\dagger$ creates a photon in an orthogonal mode (say with orthogonal polarisation) and $|\text{vac}\rangle$ is the vacuum. $\eta$ is a real parameter describing the degree of overlap of the quantum states in modes 1 and 2. Since reflection at the beam splitter requires a phase shift of $90$ degrees, we represent the BS transformation using the conventions 
\begin{align}
a_1^\dagger &\rightarrow (ia_3^\dagger+a^\dagger_4)/\sqrt{2}\nonumber,\\
a_2^\dagger &\rightarrow (a_3^\dagger+ia^\dagger_4)/\sqrt{2}
\end{align}
and similarly for the $b$ modes. The indices $3$ and $4$ denote the output ports of the BS.
Then, one has 
\begin{align}
|\psi\rangle \rightarrow \frac{1}{2}(&\eta[ia_3^\dagger a_3^\dagger+\overbrace{a_4^\dagger a_3^\dagger-a_3^\dagger a_4^\dagger}^{=\,0} + ia_4^\dagger a_4^\dagger]+ \nonumber\\
 &\sqrt{1-|\eta|^2}[ib_3^\dagger a_3^\dagger+b_4^\dagger a_3^\dagger-b_3^\dagger a_4^\dagger +ib_4^\dagger a_4^\dagger])|\text{vac}\rangle.
\end{align}
The cancellation above is a consequence of the bosonic commutation relation $[a_3^\dagger,a_4^\dagger]=0$. Now assuming we have detectors that indiscriminately register coincidences (one photon in mode 3 and one photon in mode 4), the probability of this occurring is
\begin{align}
P_c = \frac{1}{2}(1-|\eta|^2).
\end{align}
Expanding the signal and idler photons into an orthonormal time-bin basis $\langle t |t'\rangle = \delta_{tt'}$:
\begin{align}
\eta(\tau) &= \langle \text{vac}|(\eta a_1 + \sqrt{1-|\eta|^2} b_1)a^\dagger_2|\text{vac}\rangle =\langle \textrm{signal} | \textrm{idler} \rangle\nonumber\\
&= \sqrt{\alpha} \langle t |\int\int f_1^*(t-\tau_\text{sample})f_2(t'-\tau_\text{HOM})dtdt'|t'\rangle \nonumber\\
&= \sqrt{\alpha} \int f_1^*(t-\tau_\text{sample})f_2(t-\tau_\text{HOM})dt\nonumber \\
&= \sqrt{\alpha} \int f_1^*(t-\tau)f_2(t)dt,
\end{align}
where we defined $\tau=\tau_\text{sample}-\tau_\text{HOM}$. Here $f_i$ is the temporal mode function of the photon in each input port of the beam splitter. The time delay $\tau$ (which transforms $a_1$ toward $b_1$) is introduced either through a controllable translation stage, or a transparent sample of unknown refractive properties.  $\alpha$ is a positive phenomenological parameter representing residual distinguishability for perfectly synchronised modes, contributed to by polarisation, spatial mode or other mismatches as well as any imbalance in the BS. The temporal mode functions are set by the longitudinal uncertainty in the location of the downconversion event. When this is limited by the length of the crystal, the mode functions are top-hat functions, leading to a triangular dip. Here we assume the use of spectral filters which tend to broaden and reshape the temporal distribution, leading to a Gaussian dip \cite{Dauler2000}.  If $f_1=f_2$ and are both Gaussians with s.d. $\sigma/\sqrt{8}$ we then obtain 
\begin{align}
P_c =\frac{1}{2}\left(1-\alpha e^{-s^2}\right),
\label{lossless_model}
\end{align}
where we defined the normalised temporal delay $s:=\tau/\sigma$ which is a dimensionless quantity. 
\\
\\
{\emph{Loss Model:}}
Real experiments are subject to losses -- in our case they are dominated by the inefficiency of our photodetectors. We therefore model this by allowing for a photon to be lost with probability $\gamma$ immediately before detection. The full model is thus given by:
\begin{align}
\left(
\begin{array}{c}
P_0\\
P_1\\
P_2
\end{array}
\right)
=
\left(
\begin{array}{cc}
\gamma^2 & \gamma^2\\
2\gamma(1-\gamma) & 1-\gamma^2\\
1-2\gamma(1-\gamma)-\gamma^2 & 0
\end{array}
\right)
\left(
\begin{array}{c}
P_c\\
P_b
\end{array}
\right).
\end{align}
with $P_b=1-P_c$ implied by normalisation. Transforming (\ref{lossless_model}), the resultant model is:
\begin{align}
P_2&=\frac{1}{2} (1-\gamma )^2 \left(1-\alpha e^{-s^2}  \right),\\
P_1&=\frac{1}{2} (1-\gamma )^2  \left(\frac{1+3 \gamma}{1-\gamma} +\alpha  e^{-s^2}\right) ~, \\
P_0&=\gamma ^2 .
\end{align}
We have used the label $i=0,1,2$ to denote the number of detectors which click -- i.e. a total loss, bunch and coincidence respectively. The total number of incident photon pairs is given by $N=N_0+N_1+N_2$. Note that $N_0$ is a purely theoretical quantity used to define our model, and need not (and in fact cannot) be measured at all.
\\
\\
{\emph{Fisher Information:}}
The Fisher information $F_s$ is defined as a functional of a statistical model $P(i|s)$, which is a normalised set of probabilities for outcomes $i$ conditioned on the value of our target parameter $s$ (such as those above). The Fisher information in the main text may be calculated as:  
\begin{align}
F_{\tau}=\frac{1}{\sigma^2}F_s = \frac{1}{\sigma^2}\sum_i \frac{(\partial_s P(i|s))^2}{P(i|s)}.
\end{align}
\\
\\
{\emph{Maximum Likelihood Estimator:}}
 The likelihood is a multinomial distribution:
$
\mathcal{L}(N_1,N_2|\tau) \propto P_0^{N_0}P_1^{N_1}P_2^{N_2} ~,
$
(where the constant of proportionality does not depend on $s$).  We extremize the likelihood as follows
\begin{align}
0&=:(\partial_{s}\log\mathcal{L})_{\tilde{s}_{\text{MLE}}}\nonumber\\
&=\cancel{\partial_{s}(N_0\log(P_0))}\nonumber\\
&+\partial_{s}(N_1\log(P_1))_{\tilde{s}_{\text{MLE}}}+\partial_{s}(N_2\log(P_2))_{\tilde{s}_{\text{MLE}}}\nonumber\\
&=\left.\frac{N_1P_1'}{P_1}\right|_{\tilde{s}_{\text{MLE}}}+\left.\frac{N_2P_2'}{P_2}\right|_{\tilde{s}_{\text{MLE}}}\nonumber\\
&=\left.\frac{N_1P_1'}{P_1}\right|_{\tilde{s}_{\text{MLE}}}-\left.\frac{N_2P_1'}{P_2}\right|_{\tilde{s}_{\text{MLE}}}\nonumber\\
\left.N_1P_2\right|_{\tilde{s}_{\text{MLE}}}&=\left.N_2P_1\right|_{\tilde{s}_{\text{MLE}}}
\end{align}
This equation is then solved for $\tilde{s}_\text{MLE}$. We discard the minimum likelihood solution at $s=0$. The solution which corresponds to a maximum is described in the main text.  \\
\\
{\emph{Peak Information Point}:}
When $\alpha=1$, the Fisher information given in the main text is maximised near $s=0$, although it is undefined there. We have
$$
\lim_{s\rightarrow0} \left.F_s\right|_{\alpha=1}=2
$$
As $\alpha$ is lowered, two true peaks appear, moving outwards and becoming broader. When $\gamma=0$ we have 
$$
F_s = \frac{4 \alpha ^2 s^2}{e^{2 s^2}-\alpha ^2}
$$
with maximum
$$
s^*=\frac{\pm\sqrt{W\left(-\frac{\alpha ^2}{e}\right)+1}}{\sqrt{2}}
$$
for $W$ the Lambert $W$-function. 
When $\gamma\neq0$, $s^*$ can be found numerically (it has a rather weak dependence on $\gamma$). For $\alpha\rightarrow 0$, the optimum moves to the inflection point of the Gaussian, $s^*\rightarrow \pm1/\sqrt{2}$, Figure~\ref{ystar}.
\\
\\
{\emph{Bayesian analysis:}}
To avoid infinities, we use a prior distribution $p(s)$ uniform on $[-s_{\textrm{max}},s_{\textrm{max}}]$. Instead of directly maximising the likelihood, we instead use it to multiplicatively update the prior distribution before maximizing the resultant posterior distribution. This is an application of Bayes’ rule:
\begin{align}
p(s|N_1,N_2) \propto \mathcal{L}(N_1,N_2|s)p(s)
\end{align}
(again, the proportionality constant does not depend on $s$). The maximum of this posterior is unchanged (with respect to the likelihood) when the argument of the logarithm is positive. When the argument is not positive, the maximum posterior is at $s =\pm s_{\textrm{max}}$. So our full, Maximum-A-Posteriori estimator is
\begin{align}
\tilde{s}_{\textrm{MAP}} = 
\begin{cases}
      \phantom{\pm}\tilde{s}&\text{for } N_1 -  N_2(\frac{1+3\gamma}{1-\gamma }) > 0 \\
      \pm s_{\textrm{max}} &\text{for }  N_1 -  N_2(\frac{1+3\gamma}{1-\gamma }) \leq 0.
   \end{cases}
\end{align}
We set $s_{\textrm{max}} = 10$. We also have $\tilde{\tau}=\sigma\tilde{s}$ where estimators are denoted with a $\sim$.
\\
\\
{\emph{Calibration Stage:}}
Our measurement protocol comprises of several steps beginning with a full calibration of the parameters $N$, $\gamma$, $\sigma$ and $\alpha$. First, the interferometer is tuned far outside the dip and we calculate the photon loss parameter $\gamma$
\begin{align}
\tilde{\gamma} = \left. \frac{N_1-N_2}{N_1+3N_2}\right|_{s\rightarrow \infty}.
\end{align}
Next to allow us to estimate the precision of our experiment we also estimate the total number of incident photon pairs
\begin{align}
\tilde{N} = \frac{N_1+N_2}{1-\gamma^2}.
\end{align}
Now we vary $\tau_\text{HOM}$ to perform a partial scan of the dip, that covers both the $s=0$ and $s=s^*$ points.
Then we have
\begin{align}
\tilde{\alpha} = 1-\frac{2\min(N_2)}{N(\gamma-1)^2}.
\end{align}
Finally, we apply $\tilde{s}$ to the partial scan of the dip. This irons out the bell shaped dip to a roughly linear V-shape (see Appendix D). We then perform a linear fit near the target region and $\sigma$ is taken as the inverse of the gradient. We choose the size of the fitting window to be approximately 7 fs (see Appendix D). 
\\
\\
{\emph{Bias:}}
Since the dominant sources of imperfection are accounted for in our model, we expect the measured precision (related to the inverse-root of the Fisher information) and measured accuracy to closely match the theoretical quantities, although small discrepancies are to be expected due to uncontrollable sources of random and systematic error. The full CRB is 
\begin{align}
\text{Var}(\tilde{\tau})\geq \frac{(1+\frac{\partial b(\tau)}{\partial \tau})^2}{NF} \approx 
\frac{1}{NF} ~.
\label{CRBbias}
\end{align}
with $b(\tau)=\mathbb{E}(\tilde{\tau}-\tau)$ being the bias ($\mathbb{E}$ is the expected value). The Maximum Likelihood Estimator is consistent, meaning that the bias is zero in the limit of $N\rightarrow\infty$ \cite{Wald1949}.  Because we have very large $N$, the bias should therefore be negligible.

\section*{Appendix B: Experiments}
A frequency-doubled Ti:Sapphire oscillator (Coherent Chameleon Ultra II) with 130 fs duration at a repetition rate of 80 MHz is used to pump a 0.5 mm long Type II BBO crystal for wavelength degenerate Spontaneous Parametric Down Conversion (SPDC). The 808 nm signal and idler photons are spatially separated using a Polarising Beamsplitter (PBS) and then coupled into polarisation maintaining fibers where they are guided to a fiber-coupled 50:50 cube beamsplitter (HOM BS) as shown in Figure~\ref{setup_fig}. Coarse control of the interferometer delay, $\tau_\text{HOM}$, is controlled by adjusting the on-axis position of one of the fiber couplers using a translation stage (HOM Stage). In this way the HOM dip can be characterised by counting coincident events between two SPAD detectors which are positioned at the output arms of the HOM BS as the delay is changed. Timing for the coincident event detection is managed by an Event Timing Module (Picoquant Hyrdaharp 400). Fine control of the delay is achieved by moving the other fiber coupler with a piezo actuator controlled translation stage (Piezo Actuator - PI P-753.1CD). This configuration allows for precise control of the optical path length with a sub-nm resolution.

The translating wedge system is calibrated by removing the spectral filters prohibiting the SHG pump beam from reaching the BBO used for downconversion and allowing the coherent state of the laser at 808 nm to pass though the setup as a Mach-Zehnder interferometer. The beam is attenuated to the single photon level and a half waveplate before the PBS is rotated to balance the photon count level in each interferometer arm to yield high visibility interference. The period of the resulting interference fringes allows a conversion factor to be defined of a 1 $\mu$m translation of the wedges resulting in an effective path length change of approximately 17 nm.
\section*{Appendix C: Dynamic range of the Measurement Procedure}
The dynamic range may be defined as the maximum interval in $\tau$ such that the estimator is single-valued. Phase sensitive interferometers (such as the Mach-Zehnder) suffer from a `phase-wrapping' problem, where a multitude of physical time delays result in the same relative phase in the interferometer. Those physical time delays cannot then be mutually distinguished, unless some prior information is available. The dynamic range of a Mach-Zehnder interferometer is set by the wavelength of light that is used.  One can use a uniform prior distribution to exclude parameters outside a certain interval: this can restore the uniqueness of estimates, at the expense of having enough prior information (represented by the inverse width of the interval). A higher dynamic range implies that less prior information is required to avoid ambiguities. In principle the dynamic range of our Gaussian HOM interferometer is infinite because of the long tails of $f(t)$. In practice, however, one should cap the dynamic range to a region where $F$ is above some threshold value. A conservative estimate would be on the order of 1 or 2 $\sigma$, which is typically 10-100$\times$ larger than the wavelength of light in a phase dependent (e.g. MZ) interferometer.

\section*{Appendix D: Procedure for the Local Fitting of the HOM Dip}

The width of the HOM dip is the final fit parameter to be estimated. To construct an estimate, we perform a partial scan of the HOM dip which results in a list of triples ($\tau$, $N_1$, $N_2$), where $\tau$ is the `ground truth' optical delay inferred from electronic readout of the piezo stage. Using the already estimated values of $\alpha$ and $\gamma$, we reduce each triple using the estimator $\tilde{s}$ (which is nothing other than $\tilde{\tau}/\sigma$, see Eq.(3) of the main text). This estimator is a function of $N_1$ and $N_2$ and maps the list of triples into a list of pairs ($\tau,s$). This has the effect of straightening the Gaussian dip into a `vee'. Since $\tau = \sigma s$, we can perform a linear fit of these data. We chose to perform the fit in a restricted region about 7 fs wide, centred on the point of maximum Fisher information. Our estimate of $\sigma$ is simply the inverse of the gradient: $\tilde{s}=\Delta \tau / \Delta s$.
\begin{figure}
\centering
\includegraphics[width=0.6\textwidth]{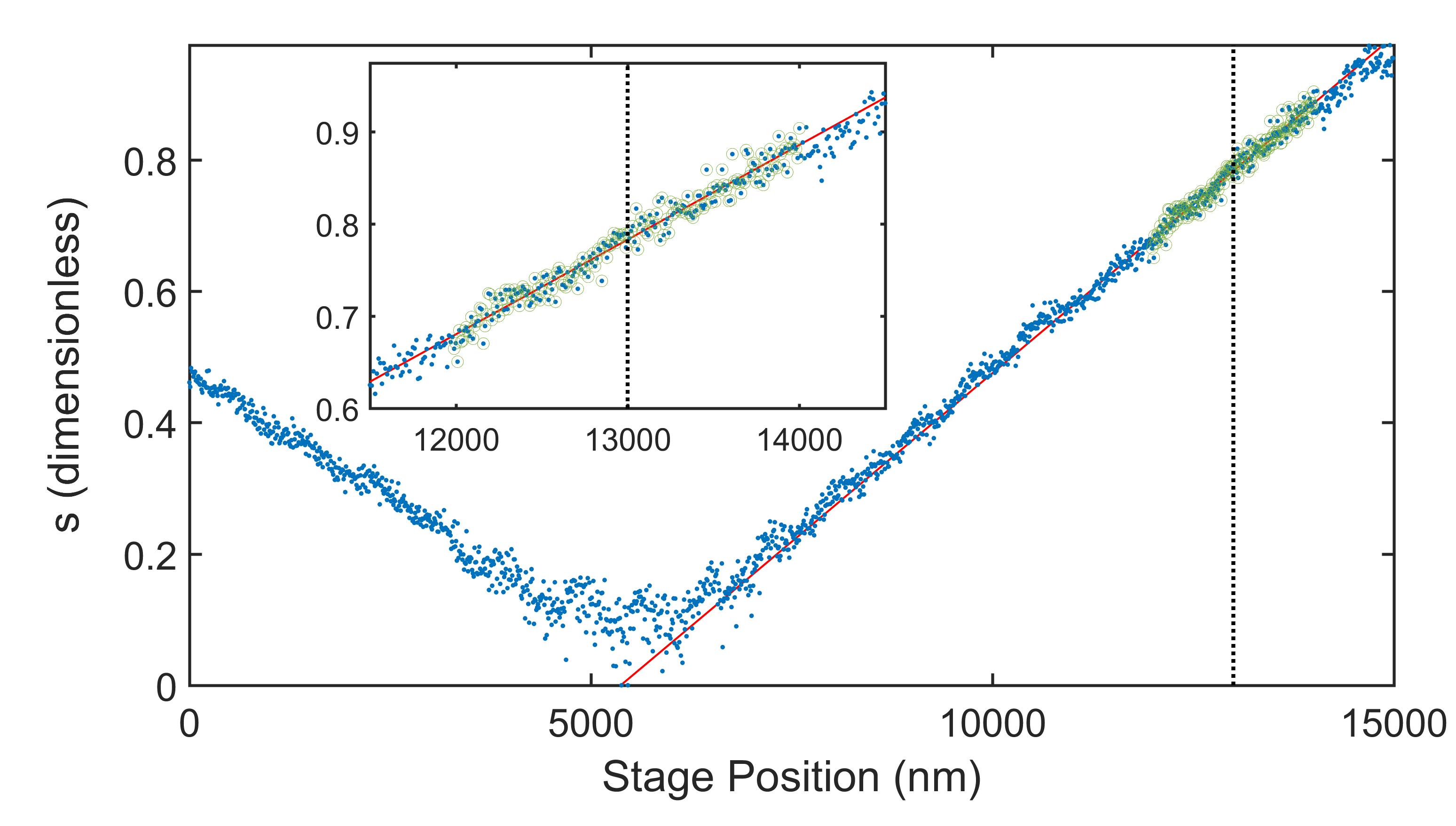}
\caption{Representative example of our fitting procedure to extract $\sigma$ during the calibration of the HOM dip.  Inset: Enlarged view of the region of interest for the sensing procedure.}
\label{all_fits}
\end{figure}

\section*{Appendix E: Activating phase fringes}
As we alluded to in the main text, it is straightforward for our setup to be converted to a phase sensitive one by including a single optical element ($\sfrac{\lambda}{2}$ waveplate before the PBS). Here we will show the full effect of an arbitrary rotation of the waveplate on the coincidence probability, and the significantly increased Fisher information that results.

\begin{figure}[t]
\centering
\includegraphics[width=0.8\textwidth]{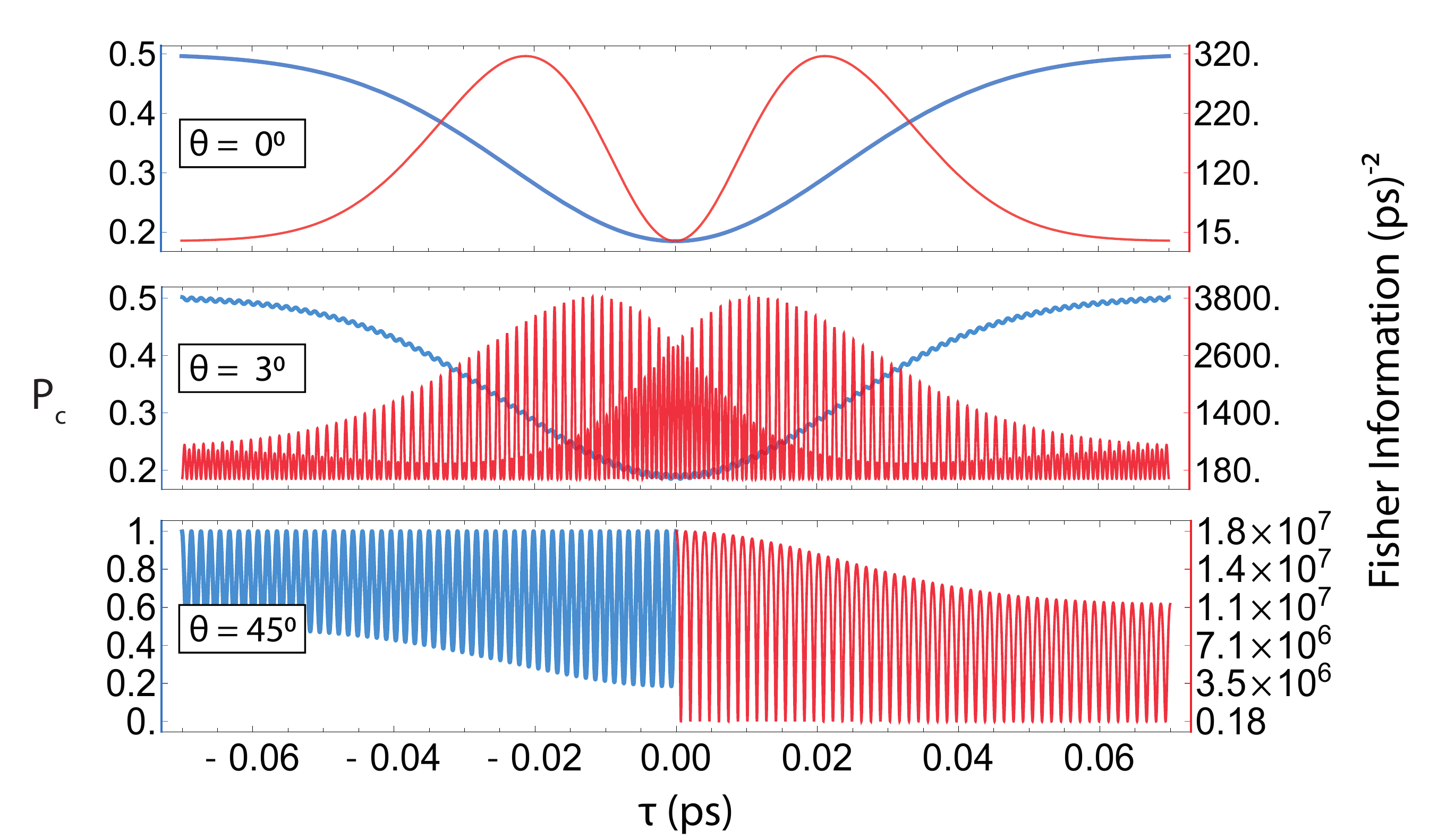}
\caption{Predicted Fisher information for a HOM with added phase-dependent fringes. The visibility of the fringes can be tuned to increase the obtainable precision albeit at the expense of introducing potential instabilities and reducing the dynamic range. These plots correspond to $\alpha=0.63$, $\sigma = 0.033$ ps, and $\nu=371$ THz (no loss is modelled, so $\gamma=0$). The curves in the lowermost panel have been alternately cutaway either side of the origin (for clarity). All curves are symmetric around $\tau=0$.
}
\label{wiggly_fish}
\end{figure}

 Consider misalignment between the downconversion crystal and the PBS. Assume that horizontal ($H$) and vertical ($V$) polarisations are defined by the PBS, and that the downconversion source is rotated around the pump propagation axis by an angle $\theta$. The true downconversion state is 
\begin{align}
|\psi\rangle_{\rm SPDC} &= \Big[ \eta (\cos(\theta) a_{1H}^\dagger + \sin(\theta) a_{1V}^\dagger)(\sin(\theta) a_{1H}^\dagger -\cos(\theta) a_{1V}^\dagger) \nonumber \\ 
& \hspace{0.5cm} + \beta(\cos(\theta) b_{1H}^\dagger + \sin(\theta) b_{1V}^\dagger)(\sin(\theta) a_{1H}^\dagger -\cos(\theta) a_{1V}^\dagger) \Big] |\text{vac} \rangle
\end{align}
where the creation operators now carry polarisation labels. The PBS has the action of transforming $V$ photons to mode 2 with a phase shift of $i$ due to reflection. It has no effect on $H$ photons:
\begin{align}
|\psi\rangle_{\rm PBS} &= \Big[ \eta (\cos(\theta) a_{1H}^\dagger + i\sin(\theta) a_{2V}^\dagger)(\sin(\theta) a_{1H}^\dagger -i\cos(\theta) a_{2V}^\dagger) \nonumber \\
& \hspace{0.5cm} + \beta(\cos(\theta) b_{1H}^\dagger + i\sin(\theta) b_{2V}^\dagger)(\sin(\theta) a_{1H}^\dagger -i\cos(\theta) a_{2V}^\dagger) \Big] |\text{vac}\rangle.
\end{align}
Next the HWP takes $V$ to $H$ and $H$ to $V$ but only in mode 2 (this should also be well aligned with the PBS):
\begin{align}
|\psi\rangle_{\rm HWP} &= \Big[ \eta (\cos(\theta) a_{1H}^\dagger + i\sin(\theta) a_{2H}^\dagger)(\sin(\theta) a_{1H}^\dagger -i\cos(\theta) a_{2H}^\dagger) \nonumber \\
& \hspace{0.5cm} + \beta(\cos(\theta) b_{1H}^\dagger + i\sin(\theta) b_{2H}^\dagger)(\sin(\theta) a_{1H}^\dagger -i\cos(\theta) a_{2H}^\dagger) \Big]  |\text{vac}\rangle \nonumber\\
&= \Big[ \eta (\cos(\theta) a_{1}^\dagger + i\sin(\theta) a_{2}^\dagger)(\sin(\theta) a_{1}^\dagger -i\cos(\theta) a_{2}^\dagger) \nonumber \\
& \hspace{0.5cm} + \beta(\cos(\theta) b_{1}^\dagger + i\sin(\theta) b_{2}^\dagger)(\sin(\theta) a_{1}^\dagger -i\cos(\theta) a_{2}^\dagger) \Big]  |\text{vac}\rangle.
\end{align}
We now drop the polarization labels since all photons have the same value now. Next we introduce the path length change of $\tau$, which gives rise to the phase shift $e^{i\phi}=e^{i2\pi \tau/\lambda}$ in mode 2. 
\begin{align}
|\psi\rangle_{\rm shift} &= \Big[ \eta (\cos(\theta) a_{1}^\dagger + ie^{i\phi}\sin(\theta)  a_{2}^\dagger)(\sin(\theta) a_{1}^\dagger -ie^{i\phi}\cos(\theta) a_{2}^\dagger) \nonumber \\
& + \beta(\cos(\theta) b_{1}^\dagger + ie^{i\phi}\sin(\theta) b_{2}^\dagger)(\sin(\theta) a_{1}^\dagger -ie^{i\phi}\cos(\theta) a_{2}^\dagger) \Big]  |\text{vac}\rangle. 
\end{align}
Now the final step is the beamsplitter (whose action is shown in the main text),
\begin{align}
|\psi\rangle_{\rm BS} &= \Big[ \frac{\eta}{2} (\cos(\theta) [ia_{3}^\dagger+a_4^\dagger] + ie^{i\phi}\sin(\theta) [a_{3}^\dagger+ia_4^\dagger]) (\sin(\theta)  [ia_{3}^\dagger+a_4^\dagger] -ie^{i\phi}\cos(\theta) [a_{3}^\dagger+ia_4^\dagger]) \nonumber \\
& \hspace{0.5cm} +\frac{ \beta}{2}(\cos(\theta) [ib_{3}^\dagger+b_4^\dagger] + ie^{i\phi}\sin(\theta) [b_{3}^\dagger+ib_4^\dagger])  (\sin(\theta) [ia_{3}^\dagger+a_4^\dagger]-ie^{i\phi}\cos(\theta)[a_{3}^\dagger+ia_4^\dagger]) \Big]  |\text{vac}\rangle  \nonumber\\
&= \Big[ \frac{\eta}{2}[a_3^\dagger(i\cos(\theta)+ie^{i\phi}\sin(\theta))+a_4^\dagger(\cos(\theta)-e^{i\phi}\sin(\theta))]  [a_3^\dagger(i\sin(\theta)-ie^{i\phi}\cos(\theta))+a_4^\dagger(\sin(\theta)+e^{i\phi}\cos(\theta))]\nonumber\\
& \hspace{0.5cm} +\frac{\beta}{2}[b_3^\dagger(i\cos(\theta)+ie^{i\phi}\sin(\theta))+b_4^\dagger(\cos(\theta)-e^{i\phi}\sin(\theta))] [a_3^\dagger(i\sin(\theta)-ie^{i\phi}\cos(\theta))+a_4^\dagger(\sin+(\theta)e^{i\phi}\cos(\theta))] \Big]  |\text{vac}\rangle.
\end{align}
The next step is to combine the correct amplitudes to get a coincidence probability. There are amplitudes for $a_3^\dagger a_4^\dagger$ and $a_4^\dagger a_3^\dagger$:
\begin{align}
\mathcal{A}_{a^\dagger_3a^\dagger_4} &= \frac{\eta}{2}\Big[ (i\cos(\theta)+ie^{i\phi}\sin(\theta))(\sin(\theta)+e^{i\phi}\cos(\theta)) \Big], \nonumber\\
\mathcal{A}_{a^\dagger_4a^\dagger_3} &= \frac{\eta}{2}\Big[ (\cos(\theta)-e^{i\phi}\sin(\theta))(i\sin(\theta)-ie^{i\phi}\cos(\theta)) \Big].
\end{align}
These operators commute, because of the orthogonal spatial modes: therefore $a_3^\dagger a_4^\dagger=a_4^\dagger a_3^\dagger$ and we should add these terms coherently, allowing the HOM interference effect to operate. There are two more amplitudes (that cannot interfere) for $b_3^\dagger a_4^\dagger$ and $b_4^\dagger a_3^\dagger$ (where once again operators commute):
\begin{align}
\mathcal{A}_{b^\dagger_3a^\dagger_4} &= \frac{\beta}{2}\Big[(i\cos(\theta)+ie^{i\phi}z\sin(\theta))(\sin(\theta)+e^{i\phi}\cos(\theta)) \Big],\nonumber\\
\mathcal{A}_{b^\dagger_4a^\dagger_3} &= \frac{\eta}{2}\Big[(\cos(\theta)-e^{i\phi}\sin(\theta))(i\sin(\theta)-ie^{i\phi}\cos(\theta)) \Big].
\end{align}
Computing these terms and making the substitutions  $\beta^2=1-\eta^2$,  $\eta^2\rightarrow \alpha e^{-\tau^2/\sigma^2}$ and $\phi\rightarrow 2\pi c\tau/\lambda = 2\pi \tau\nu$ (where $\nu$ is the frequency of the light), we have 
\begin{align}
P_c &= |\mathcal{A}_{a^\dagger_3a^\dagger_4}+\mathcal{A}_{a^\dagger_4a^\dagger_3}|^2+|\mathcal{A}_{b^\dagger_3a^\dagger_4}|^2+|\mathcal{A}_{b^\dagger_4a^\dagger_3} |^2\nonumber\\
&=\frac{1}{2} \left(\sin ^2(2 \theta ) \left(\alpha  e^{-\frac{\tau ^2}{\sigma ^2}}+1\right) \cos ^2(2 \pi  \nu  \tau )-\alpha  e^{-\frac{\tau ^2}{\sigma ^2}}+1\right).
\end{align}
Due to the the HWP erasing the which-path information, the interferometer can then act like a Mach Zehnder, and each photon independently oscillates between SPAD1 and SPAD2. There will be a fixed phase offset of 90 degrees between the two photons, because they began life orthogonal to each other. This means two independent oscillations in antiphase, each at the frequency of the signal/idler. This results in a coincidence probability oscillating at twice that frequency. This is $N=2$ N00N interferometry, which is easily switched on in our setup -- this could provide higher information and a way to remove ambiguities whilst not sacrificing the dynamic range of the HOM setup. We observe that it is a second order effect that recovers the usual HOM dip when $\theta\rightarrow0$ (see Figure~\ref{wiggly_fish}.
The full Fisher information is 
\begin{align}
F = -\frac{4 \left(\pi  \nu  \sigma ^2 \sin ^2(2 \theta ) \left(\alpha +e^{\frac{\tau ^2}{\sigma ^2}}\right) \sin (4 \pi  \nu  \tau )+\alpha  \tau  \sin ^2(2 \theta ) \cos ^2(2 \pi  \nu  \tau )-\alpha  \tau \right)^2}{\sigma ^4 \left(\alpha +e^{\frac{\tau ^2}{\sigma ^2}}\right) \left(\sin ^2(2 \theta ) \cos ^2(2 \pi  \nu  \tau )-1\right) \left(\sin ^2(2 \theta ) \left(\alpha +e^{\frac{\tau ^2}{\sigma ^2}}\right) \cos ^2(2 \pi  \nu  \tau )-\alpha +e^{\frac{\tau ^2}{\sigma ^2}}\right)}.
\label{long_eq}
\end{align}

In the case of $\theta\rightarrow 0$, this expression reduces to the lossless version of the Fisher information used in the main paper. In the case of $\theta\rightarrow 45^{\circ}$, it simplifies to 
\begin{align}
F = \frac{8 \left(\alpha  \tau  \sin (2 \pi  \nu  \tau )-2 \pi  \nu  \sigma ^2 \left(\alpha +e^{\frac{\tau ^2}{\sigma ^2}}\right) \cos (2 \pi  \nu  \tau )\right)^2}{\sigma ^4 \left(\alpha +e^{\frac{\tau ^2}{\sigma ^2}}\right) \left(\alpha  (\cos (4 \pi  \nu  \tau )-1)+e^{\frac{\tau ^2}{\sigma ^2}} (\cos (4 \pi  \nu  \tau )+3)\right)}.
\end{align}
Setting $\alpha=0.63$, $\sigma = 0.033$ ps, $\nu=371$ THz, we find that the peak Fisher information for $\theta=45^{\circ}$ is approximately 24000 times higher that for $\theta=0^{\circ}$. This equates to a potential 155 fold improvement in precision.

\section*{Appendix F: List of Fitting Parameters \& Results}

The table below shows the collection of parameters used to analyse the data shown in all figures along with the final results.
\begin{table}[h]
\vspace{0.5cm}

\begin{tabular}{c||r|r|r|r|r|||r|r|r|r||r||r|r|r}
&&\multicolumn{4}{|c|||}{$\delta \tau$ (attosecond)}&\multicolumn{4}{|c||}{$c\delta\tau/n$ (nanometre)}&\multicolumn{4}{c}{fitted parameters} \\ \hline
Figure & $n$ & expected &  measured  & accuracy & precision  & expected & measured  & accuracy & precision & $N$ &  $\sigma$ (ps)  &  $\gamma$  &  $\alpha$ \\\hline
\hline
2 & 1.0 & - & scan & - & - & - & scan & - & - & 3.72e+5 & 0.07 & 0.88 & 0.53 \\
2 & 1.0 & - & scan & - & - & - & scan & - & - & 3.72e+5 & 0.07 & 0.87 & 0.78 \\
2 & 1.0 & - & scan & - & - & - & scan & - & - & 3.72e+5 & 0.06 & 0.87 & 0.92 \\

\hline
3 and 4 & 1.0 & -33.33 & -28.97 & -4.36 & 14.00 & -10.00 & -8.69 & -1.31 & 4.20 & 434 bn & 0.03 & 0.87 & 0.63 \\
4 & 1.0 & -66.67 & -57.20 & -9.47 & 46.18 & -20.00 & -17.16 & -2.84 & 13.85 & 276 bn & 0.05 & 0.87 & 0.65 \\
4 & 1.0 & -100.00 & -96.27 & -3.73 & 15.48 & -30.00 & -28.88 & -1.12 & 4.64 & 287 bn & 0.04 & 0.86 & 0.69 \\
4 & 1.0 & -6.67 & -6.16 & \textbf{-0.51} & 13.41 & -2.00 & -1.85 & -0.15 & 4.02 & 190 bn & 0.03 & 0.87 & 0.63 \\
4 & 1.0 & -50.00 & -60.84 & 10.85 & 30.29 & -15.00 & -18.25 & 3.25 & 9.09 & 63 bn & 0.03 & 0.89 & 0.73 \\
\hline
4 (wedges) & 1.5 & -56.67 & -69.36 & 12.69 & \textbf{4.72} & -11.33 & -13.87 & -2.18 & 0.94 & 414 bn & 0.03 & 0.88 & 0.75 \\
\hline
4 & 1.0 & -16.67 & -17.47 & 0.80 & 6.19 & -5.00 & -5.24 & 0.24 & 1.86 & 361 bn & 0.03 & 0.87 & 0.65 \\
%
%
4 & 1.0 & -6.67 & -16.68 & 10.02 & 7.41 & -2.00 & -5.01 & 3.01 & 2.22 & 148 bn & 0.03 & 0.87 & 0.63 \\
4 & 1.0 & -1.67 & -9.43 & 7.77 & 7.49 & -0.50 & -2.83 & 2.33 & 2.25 & 321 bn & 0.02 & 0.87 & 0.69 \\
4 & 1.0 & -1.67 & -6.36 & 4.69 & 25.13 & -0.50 & -1.91 & 1.41 & 7.54 & 502 bn & 0.05 & 0.89 & 0.68 \\
\hline\hline 
average & - & - & - & \textbf{6.48} & \textbf{17.03} & - & - & \textbf{1.78} & \textbf{5.06} & - & -& - & - \\
\end{tabular}
\caption{Summary of all measurements and parameters used in the fitting procedure. The final row corresponds to an average of the absolute value of each column. The factor $1\times 10^ 9$ is denoted bn for billions. $n=1$ or $1.5$ is the refractive index of the sample material (air or glass respectively). The single datapoint relating to the glass wedges is highlighted with horizontal lines. Average and best case values are shown in bold face.}
\end{table}

\end{document}